# "Implementación de una fórmula analítica para el calculo del coeficiente M (3000)F2"

_Arian Ojeda_ [(1)], **Silvio González** [(1)], **Katy Alazo** [(1)], **Alexander Calzadilla.**

*( 1) Instituto de Geofísica y Astronomía, Calle 212 No. 2906 e/ 29 y 31, La Coronela, La Lisa, Ciudad de La Habana, CP 11 600, Cuba. C. Eléct.: arian@iga.cu*

**RESUMEN**

Es de gran importancia la determinación del coeficiente M(3000)F2 en el análisis de ionogramas de la estación de sondeo vertical de la Ionosfera. Este coeficiente se obtiene de dividir la frecuencia máxima utilizable (MUF) para una distancia de 3000 Km entre la frecuencia crítica de la capa F2 (foF2). Para determinar el M(3000)F2 actualmente se usa un método gráfico en nuestra estación de sondeo vertical de la Ionosfera. El objetivo de este trabajo es implementar un método analítico que permita obtener directamente el coeficiente M(3000)F2. De esta forma podrá ser programado e incorporado como parte del proceso de elaboración de los ionogramas en la Estación Habana. Ahora con el uso de un ordenador al poder determinar varios puntos del Ionograma, se pueden usar estos (f; h') para calcular de forma analítica el M(3000)F2. La comparación del método analítico implementado con el método grafico muestra que el primero es mas exacto, y se disminuye los errores a la hora de determinar el M(3000)F2.

**ABSTRACT**

Determining the M(3000)F2 coefficient is very important for ionograms analysis made by the station for the Ionospheric vertical scanning. This coefficient is the result of the maximum usable frequency (MUF), for to 3000 km distance, divided by the critical frequency for the F2 layer (FoF2). Nowadays, the graphic method to determine the M(3000)F2 coefficient is being used in our station for the Ionospheric vertical scanning. The purpose is to implement an analytic method that allows us the direct obtaining of M(3000)F2 coefficient so it could be programmed and incorporated as part of ionograms elaboration process in Havana station. Using a PC various points in a ionogram can be determined. This data (f; h') are used to analytically calculate the M(3000)F2 coefficient. Comparison between the implemented analytic method and the graphic one shows that the former is more accurate and errors are diminished in M(3000)F2 coefficient determination.

## Introducción

En Enero de 2003 se concluye la construcción de una tarjeta con un conversor análogo digital (A/D) incorporado para Bus ISA de una PC compatible IBM, y sus correspondientes programas para la captación, almacenamiento y visualización de los ionogramas en el monitor de la PC. Además del control de algunas operaciones de la Estación Ionosférica. Ahora la elaboración de los ionogramas se realiza en el monitor de la propia PC. Obteniéndose fácilmente los distintos parámetros que se toman del ionagrama para llenar las tablas diarias.

A raíz de esto surgió el problema de cómo facilitar la obtención del coeficiente M(3000)F2, pues para su determinación se utilizaba un método grafico que no era practico a la hora de simplificar el trabajo. Pues se hace necesario tomar tres puntos del ionograma y representarlos en un grafico semilogaritmico donde aparecen distintas curvas de transmisión [1] que corresponden a distintas frecuencias oblicuas. De éste grafico se selecciona la curva de transmisión a la cual la recta que forman esos tres puntos es tangente.

La relación existente entre la frecuencia de incidencia inclinada y la frecuencia de sondeo vertical de la ionosfera se puede determinar a partir de tres teoremas:
1- Ley de la secante.
2- Teorema de Breit y Tuve.
3- Teorema de Martyn.

El inconveniente que tienen esto tres teoremas es que consideran la tierra y la ionosfera plana, sin campo magnético [2]. Por tanto se hace necesario corregir la expresión que de ello se derive.

En la figura 1 superior se muestra la relación existente entre la altura y la frecuencia cuando el transmisor y el receptor se encuentran en el mismo punto, o sea, sondeo vertical de la ionosfera. En las graficas que vienen abajo se muestra que si separamos el receptor del transmisor una distancia determinada, la frecuencia aumenta hasta un valor máximo. Debido a la curvatura de la tierra, esta relación se cumple hasta 4000 Km. aproximadamente, suponiendo la altura de reflexión a 300 Km.

En la figura 2 se puede ver la relación existente entre la frecuencia de sondeo vertical y la frecuencia oblicua. A partir del ionograma de sondeo vertical se obtiene el ionograma de sondeo oblicuo. A medida que aumenta la frecuencia vertical aumenta la frecuencia oblicua hasta un punto máximo de donde comienza a disminuir. Esta relación entre ambas frecuencias se obtiene de relacionar la expresión correspondiente al teorema de Martyn y la ley de la secante a una distancia de 3000 Km. Las ecuaciones que se obtienen se muestran a continuación:

La ley de la secante es:

$$f_{obl} = f_v \sec\phi_0 \rightarrow donde \rightarrow \sec\phi_0 = \sqrt{\frac{D^2}{4} + h^2}\,\frac{1}{h} \qquad (1)$$

Al obtener esta formula no se puede hallar $f_{obl}$ debido a que mediante el sondeo vertical no se puede obtener una altura real de reflexión de la frecuencia $f_v$.

Del teorema de Martyn se plantea que si la frecuencia de incidencia vertical $f_v$ y la frecuencia de incidencia oblicua $f_{obl}$, se reflejan de una misma altura real, la altura virtual de incidencia vertical de la frecuencia $f_v$ es igual a la altura de la trayectoria triangular equivalente de la frecuencia de incidencia oblicua $f_{obl}$.

$$h'_{fv} = h'_{fobl} \qquad (2)$$

Por tanto si transformamos la expresión 1 obtenemos:

$$f_{obl} = fo\sqrt{1 + \left(\frac{D}{2h'}\right)^2}\text{, donde D = 3000 Km.} \qquad (3)$$

Esta es la expresión usada en la figura 2, ahora debido a lo que plantea el teorema de Martyn sí podemos usar la altura virtual de reflexión de la frecuencia $f_v$ para calcular $f_{obl}$. Pero no puede ser usada directamente para calcular $f_{obl}$ pues en ella no se ha tomado en cuenta la influencia del campo magnético, la curvatura de la tierra y de la Ionosfera. La teoría general demuestra, que el no tener en cuenta el campo magnético de la tierra, conduce a graves errores en el MUF para cortas distancias; con el aumento de la distancia los errores disminuyen hasta algunos por cientos solamente. Por otro lado, con el aumento de la distancia, se notan aún más las curvaturas de la tierra e ionosfera.

El objetivo de nuestro trabajo consiste en buscar una expresión que tenga en uno de sus términos la influencia que ejerce la curvatura de la ionosfera. Con esta expresión se podrá calcular las frecuencias oblicuas correspondientes a cada par de puntos de frecuencia de sondeo vertical, altura virtual, ($f_0$, h'). La máxima frecuencia oblicua que se obtenga es la que se llama **frecuencia máxima utilizable** (de ahora en adelante **MUF**) y que dividida por la frecuencia crítica de la capa F2 (foF2) da el buscado coeficiente M(3000.)F2. Así esta metodología podrá ser usada en la elaboración de un Software que permita hacer el calculo del M(3000)F2. Esto dará mayor rapidez y precisión al analizar el ionograma en nuestra estación de sondeo vertical de la ionosfera.

## Materiales Y Métodos

Para generar las curvas de transmisión (que corresponden a cada frecuencia oblicua) se usan los siguientes coeficientes que fueron tomados de [3]. Estos coeficientes tienen corregido la acción de la curvatura de la ionosfera.

| Altura virtual | 200 | 250 | 300 | 350 | 400 | 500 | 600 | 700 | 800 |
|---|---|---|---|---|---|---|---|---|---|
| M(3000) | 4.55 | 4.05 | 3.65 | 3.33 | 3.08 | 2.69 | 2.40 | 2.20 | 2.04 |

Para graficar cada curva se calculó el valor de fo para cada altura, y luego esos pares de puntos se plotearon en un grafico de fo vs. h', para cada MUF. Las curvas MUF ploteadas son las siguientes:

{2.5, 3., 3.5, 4., 4.5, 5., 6, 7, 8, 9, 10, 11, 12, 13, 14, 15, 16, 17, 18, 19, 20, 22, 24, 26, 28, 30, 32, 34, 36, 38, 40, 45, 50, 55, 60, 70, 80}

**Tabla.1-** Valores de MUF graficados.

La forma de obtener fo es:

fo = MUF / M(3000)

Por tanto para cada MUF obtengo 9 puntos correspondientes a cada altura que se pueden interpolar y obtener la curva de transmisión. En este caso se obtienen 37 curvas de transmisión (ver figura 3). Estas curvas son las que se usan en el calculo del M(3000)F2 por el método grafico.

A partir de (3) se obtiene la expresión que relaciona, la altura virtual con las frecuencias oblicuas (MUF) y frecuencias verticales (fo):

$$h' = \frac{D}{2\sqrt{\left(\frac{MUF}{fo}\right)^2 - 1}} \ ; \qquad (4) \text{, donde D = 3000 Km.}$$

Si graficamos esta función con la esperanza de obtener las curvas de nivel vemos que nos dan diferentes a la de la figura 3 (ver figura 4). Esto se debe a la influencia que ejerce la curvatura de la ionosfera. Todo esto se corrige con un coeficiente k [4] que se coloca en la ley de la secante como k Sec (φ). De esta manera la ecuación (4) quedaría:

$$h' = \frac{D}{2\sqrt{\left(\frac{MUF}{kfo}\right)^2 - 1}} \qquad (5) \text{ donde D = 3000 Km.}$$

De aquí:

$$h' = \frac{D}{2\sqrt{\left[\frac{M(3000)}{k}\right]^2 - 1}} \qquad (6)$$

Los valores que toma este coeficiente k se sabe que está implícito en los coeficientes M(3000) de la tabla 1. Luego usando la expresión 6 podemos calcular el valor de k a cada altura.

$$k = \frac{M(3000)}{\sqrt{\left(\frac{D}{2h'}\right)^2 + 1}} \qquad (7)$$

Estos valores fueron calculados para cada altura y dieron:

| h' | k |
|---|---|
| 200 | 0.601345 |
| 250 | 0.665816 |
| 300 | 0.715824 |
| 350 | 0.756675 |
| 400 | 0.793601 |
| 500 | 0.850653 |
| 600 | 0.891338 |
| 700 | 0.930348 |
| 800 | 0.96 |

**Tabla.2**- Valores del parámetro de corrección k obtenida para cada altura a partir de la ecuación 7.
En la figura 5 se muestran graficados estos valores.

Por tanto si usamos el método analítico para determinar el MUF (dado por la expresión 5) es necesario ubicar cada par de puntos de (frecuencia, altura); en el rango de altura que se muestran en la tabla 3: y calcular el valor de k de la ecuación de la recta de interpolación entre los puntos dados.

| Rango de altura | Ecuación de k |
|---|---|
| 200 y 250 | k = 0.00129 h + 0.34346 |
| 250 y 300 | k = 0.001 h + 0.41578 |
| 300 y 350 | k = 8.1702*$10^{-4}$ h + 0.47072 |
| 350 y 400 | k = 7.3852*$10^{-4}$ h + 0.49819 |
| 400 y 500 | k = 5.7052*$10^{-4}$ h + 0.56539 |
| 500 y 600 | k = 4.0685*$10^{-4}$ h + 0.64723 |
| 600 y 700 | k = 3.901*$10^{-4}$ h + 0.65728 |
| 700 y 800 | k = 2.9652*$10^{-4}$ h + 0.72278 |

**Tabla.3**- Rango de alturas para las cuales podemos obtener el coeficiente de corrección k que se usa en la ecuación 8 para obtener las distintas frecuencias oblicuas.

Posteriormente se sustituye éste valor de k, h' y fo en la ecuación 5 con MUF despejado (ver ecuación 8). Esto se hace para cada par de puntos del ionograma y finalmente se toma el valor máximo de esta operación que será el MUF escogido para obtener el M(3000)F2 al dividirlo por la frecuencia crítica de la capa F2 ver ecuación 9

.

$$MUF = kfo\sqrt{1+\left(\frac{D}{2h'}\right)^2} \qquad (8)$$

$$M(3000)F2 = \frac{MUF_{Maximo}}{foF2} \qquad (9)$$

Todos los calculas y gráficos realizados se efectuaron con ayuda del paquete de programas Mathematic versión 5.0. [5].

## Resultados Y Discusión

Finalmente la formula de trabajo a la que hemos llegado está dada por la expresión 8 teniendo en cuenta que el valor de k que se toma está dado por una función lineal de la altura virtual. La ecuación de k que se tome depende del rango de altura en que esté la altura virtual (ver tabla 3). La Forma de proceder a la hora de usar este método es el siguiente:

Si se toman tres puntos del ionograma con un orden determinado o sea de menor a mayor frecuencia, para cada punto se calcula la frecuencia oblicua dada por la expresión 8. Luego el valor de la frecuencia oblicua calculada debe ir aumentando hasta un valor máximo a partir del cual comienza a disminuir. Luego el valor máximo es el que se toma como MUF.

Para comprobar la efectividad de este método se tomaron en cada una de las 24 horas exactas del día 5 de noviembre de 2003 tres pares de puntos de altura frecuencia. Los que generalmente se usan para determinar la MUF con el método grafico. Además se tomó el valor de MUF reportado. Los valores calculados por el método grafico y el método implementado por nosotros se muestran en la tabla 4. En dicha tabla se muestran la diferencia que existen entre ambos valores reportados. Esta diferencia no excede en más de 0.4.

En el final del anexo se muestran todos los tríos de puntos de altura frecuencia que se toman para calcular la MUF por el método gráfico. Además se puede ver el valor de MUF que se puede obtener de representar dichos puntos en el grafico de las curvas de transmisión. Al lado de cada par de puntos se muestra el valor de MUF que se obtiene de sustituir este par de puntos en la formula 8 implementada por nosotros, con su respectivo valor de k que ya hemos explicado.

De las 00:00 a las 2:00 se tomaron tres pares de puntos que estuvieran arriba de los que se toman normalmente. Esto se realizó para comprobar que en ese intervalo los valores de frecuencia oblicua decrecían. Como se puede apreciar en los tres casos analizados se muestra esto. De las 8:00 a las 10:00 se hizo lo contrario, pues se tomaron además de los puntos que se usan normalmente para calcular el MUF, tres puntos que estuvieran por debajo. Esto se realizó para comprobar que en esté intervalo debía existir un crecimiento de los valores de frecuencia oblicua calculada con la ecuación 8. Como se puede ver al final del anexo este resultado se obtuvo sin contratiempos.

De lo visto anteriormente podemos concluir que como se plantea en la teoría. La formula implementada por nosotros muestra que los valores de frecuencia oblicua crecen hasta un máximo para después decrecer. Analicemos ahora qué ocurre en el entorno de los puntos que se toman para implementar el método grafico. Se puede ver que de las 24 horas en 15 el segundo punto que se toma para implementar el método grafico da un valor máximo de frecuencia oblicua, y por tanto este es el que reportamos por nuestro método como frecuencia máxima utilizable (MUF). En 8 casos el valor máximo dio en el último punto. Y en 1 caso dio por arriba. En estos casos se reportó este valor por ser el máximo, pero no estamos conformes con este resultado porque aquí no se ha ubicado el máximo con precisión.

**¿Qué se propone para programar un software que solucione esté problema?**

Se propone que al analizar el Ionograma seleccionen tres pares de puntos que el programa incluirá automáticamente. O sea que tomará el primer punto y calculará el valor de frecuencia oblicua MUF1, luego el segundo (sea este valor MUF2) y por último el tercero (Sea este valor MUF3). Aquí puede suceder que:

1- Si MUF2 < MUF1 ocurre que se tomaron valores por encima del buscado y por tanto debe aparecer un mensaje en pantalla que oriente a tomar tres nuevos valores menores que MUF1. Por tanto se comienza a realizar el análisis. Esto nos ocurrió a la 1:00 en el día analizado.

2- Si MUF3 > MUF2 > MUF1 ocurre que se tomaron valores por debajo del máximo y por tanto debe aparecer un mensaje en pantalla que oriente a tomar tres nuevos valores mayores que MUF3. Por tanto se comienza a realizar el análisis. Esto nos ocurrió a las 3:00, a las 8:00, a las 9:00, a las 10:00, a las 13:00, a las 14:00, a las 21:00 y a las 23:00. En todos los casos nos quedamos con el último valor pero a la hora de implementar el software es obligatorio tomar tres nuevos puntos para ser más exactos en nuestros resultados.

3- Si MUF2 > (MUF1 y MUF3) entonces es el valor buscado y se reporta como frecuencia máxima utilizable.

Es de destacar que cuantos mas cercanos se tomen los puntos, mejor será el valor reportado de MUF. Se hace necesario fijar un valor para el cual la diferencia entre las MUF no debe ser mayor, así evitaríamos que si se toman valores muy distantes pero que cumplan con la condición 3 se seleccione MUF2 como frecuencia máxima utilizable. Pues evidentemente en este caso estaríamos cometiendo un error de exactitud grande.

**¿Errores que se evitan?**

Al implementar el método grafico se cometen errores de escala a la hora de representar los puntos en la curva de transmisión, Además de errores de lectura al tomar los tres mejores puntos que den la recta tangente a la curva de transmisión de la MUF.

Con este nuevo método evitamos los errores de escala pues el calculo se realiza mediante una formula, Además que podemos minimizar los errores de lectura pues el programa que se implementará te dirá si debes escoger otros puntos. El principal error que tiene la formula está dada por el error de la pendiente al calcular k.

Por tanto la exactitud del valor reportado va a ser mayor en este método que en el método grafico.

Es bueno destacar que si pudiéramos seleccionar mas de tres puntos del Ionograma y que estuvieran lo mas próximo posible sería mas conveniente para realizar el calculo de la frecuencia máxima utilizable. Por tal motivo se propone que las frecuencias que se tomen no deben diferenciarse en más de 0.2 MHz. De ser posible en no más de 0.1 MHz.

## Conclusiones

Parta el calculo del M(3000)F2 se ha estado usando durante años el método grafico producto de los errores que produce el coeficiente k. Al no tener un ordenador que procese los pares de puntos del Ionograma, se hace necesario superponer el grafico h'f sobre las curvas de transmisión. De esa forma al tomar tres pares de puntos del Ionograma en la región cercana a la frecuencia crítica de la capa y ponerlo en una escala semilogaritmica, lo estamos linealizando. Así cuando se encontraba una recta paralela a estas curvas de transmisión. Estábamos encontrando el máximo valor de la frecuencia oblicua donde el grafico h'f intercepta en un solo punto a la curva de transmisión. O sea la frecuencia máxima utilizable o MUF.

Ahora con el uso de un ordenador al poder determinar varios puntos del Ionograma, se pueden usar los pares de puntos h'f para calcular de forma analítica con el método explicado anteriormente el M(3000)F2.

La comparación del método analítico implementado con el método grafico muestra que el primero es mas exacto, y disminuimos los errores a la hora de determinar el M(3000)F2.

**Referencias Bibliográficas**

1- SMITH, N. (1939): The relation of radio sky-wave transmission to ionosphere measurements, Proc . IRE, 27, p. 332.

2- J.W. Wright, R. W. Knecht, K Davies, (Aug, 1956), Manual On Ionospheric Vertical Soundings For The International Geophysical Year, Prepared at CRPL National Bureau of Standards Boulder, Colorado, USA. P , 67 .. 76.

3-W. R. Piggott, K. Rawer, (second edition November, 1972), World Data Center A For Solar – Terrestrial Physics, Report UAG-23. U.R.S.I, Hand Book of Ionogram Interpretation and Reduction, Second Edition,

4- Kenneth Davies (1989), Ionospheric Radio, Space Environment Laboratory Boulder, Colorado, p.171 – 172.

5- Wolfram Research, www.wolfram.com mathematica 5.0, (2003).

## Anexos

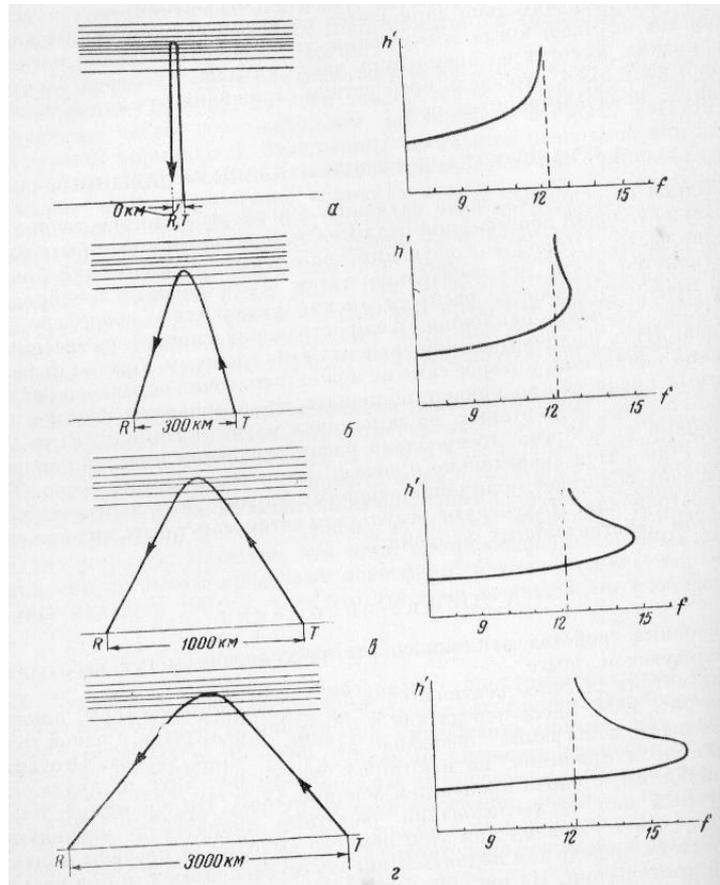

**Fig.1-** Cambio en la forma de la curva h'f a medidas que la distancia entre el transmisor y el receptor es incrementada.

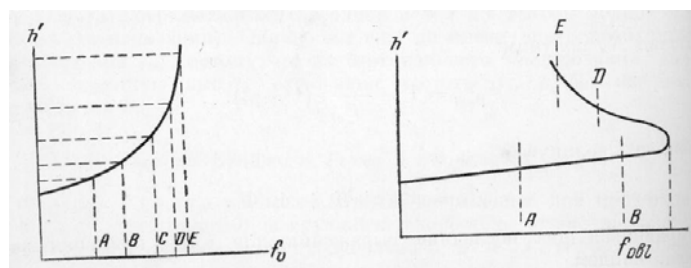

**Fig.2-** Con un par de puntos de altura virtual - frecuencia de sondeo vertical del ionograma que se encuentra en la grafica de la izquierda, podemos a partir del teorema de Martyn y la ley de la secante obtener el ionograma de sondeo inclinado para una distancia de 3000 Km. Vemos que existe un valor de frecuencia máxima oblicua (MUF) cuando se toman los valores correspondientes al punto c en el ionograma de incidencia oblicua.

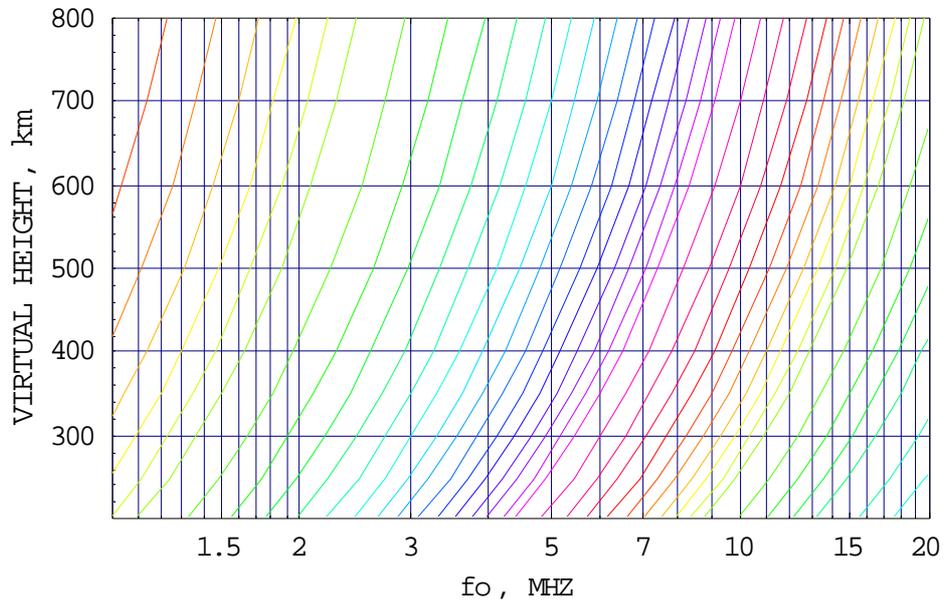

**Fig.3**- Curvas de transmisión generadas para una distancia de 3000 km. Se muestran las curvas correspondientes a las frecuencias oblicuas: 2.5, 3., 3.5, 4., 4.5, 5., 6, 7, 8, 9, 10, 11, 12, 13, 14, 15, 16, 17, 18, 19, 20, 22, 24, 26, 28, 30, 32, 34, 36, 38, 40, 45, 50, 55, 60, 70, 80, de izquierda a derecha.

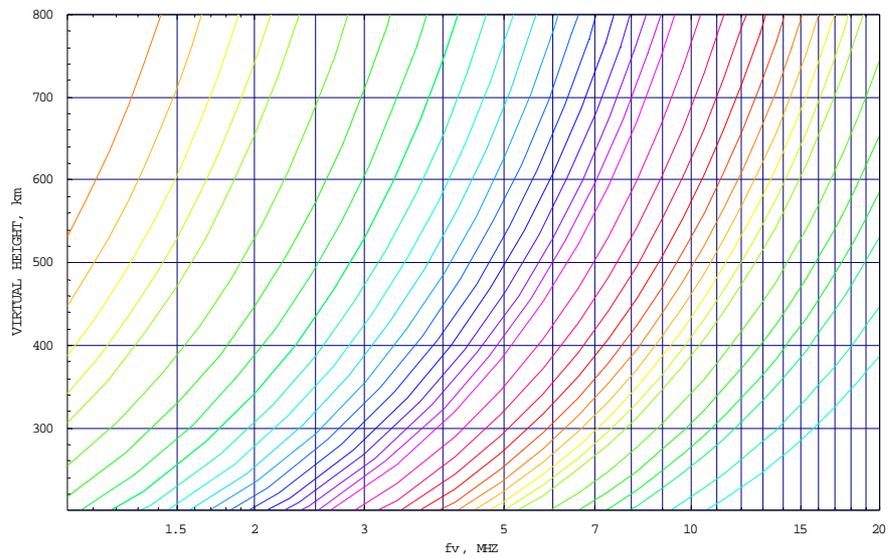

**Fig.4**- Curvas de transmisión generadas a partir de la ecuación 4 , que la no tener el coeficiente de corrección k, no da curvas de transmisión como la de la figura 3.

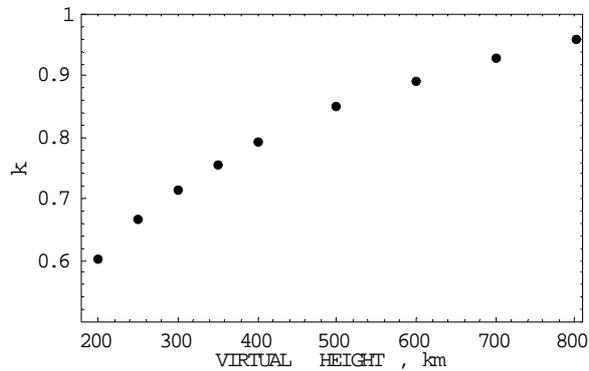

**Fig.5**- Relación entre el coeficiente de corrección k y las distintas alturas virtuales, para su calculo se usó la formula 7.

| Hora | Método grafico | Método implementado | Diferencia |
|------|----------------|---------------------|------------|
| 00 | 14.8 | 15.08 | 0.28 |
| 01 | 14.9 | 15.03 | 0.13 |
| 02 | 14.9 | 14.69 | 0.21 |
| 03 | 14.3 | 14.41 | 0.11 |
| 04 | 14.2 | 14.13 | 0.07 |
| 05 | 13.8 | 13.66 | 0.14 |
| 06 | 11.8 | 11.69 | 0.11 |
| 07 | 21.0 | 20.85 | 0.15 |
| 08 | 28.3 | 28.26 | **0.04** |
| 09 | 29.9 | 29.4 | **0.50** |
| 10 | 31.5 | 31.15 | 0.35 |
| 11 | 31.6 | 31.39 | 0.21 |
| 12 | 29.0 | 28.78 | 0.22 |
| 13 | 31.0 | 30.60 | **0.40** |
| 14 | 32.0 | 31.90 | 0.10 |
| 15 | 33.9 | 33.86 | **0.04** |
| 16 | 37.0 | 36.84 | 0.16 |
| 17 | 36.0 | 36.44 | 0.44 |
| 18 | 28.0 | 27.89 | 0.11 |
| 19 | 17.5 | 17.44 | 0.06 |
| 20 | 12.2 | 12.03 | 0.17 |
| 21 | 14.5 | 14.27 | 0.23 |
| 22 | 14.6 | 14.53 | 0.07 |
| 23 | 16.4 | 16.18 | 0.22 |

**Tabla. 4**- Comparación entre el método grafico y el método implementado. Andemos se puede apreciar la diferencia entre los resultados. En negritas se muestran los dos mayores diferencias y la dos menores. Todos estos valores corresponden al día 5 de noviembre de 2003.

### Día 5 Noviembre del 2003

*Las 00:00*
fo = 4.0 y h = 310 MUF = 15,03;
fo = 4.2 y h = 325 MUF = **15,08**;
fo = 4.5 y h = 355 MUF = 14,857; Con el método grafico se obtuvo 14.8
Por arriba
fo = 4.6 y h = 375 MUF = 14,7;
fo = 4.7 y h = 395 MUF = 14.58;
fo = 4.8 y h = 435 MUF = 14.02;

*Las 01:00*
fo = 4.0 y h = 310 MUF = **15.03**;
fo = 4.4 y h = 345 MUF = 14.77;
fo = 4.6 y h = 370 MUF = 14.82; Con el método grafico se obtuvo 14.9
Por arriba
fo = 4.6 y h = 370 MUF =14.82 ;
fo = 4.7 y h = 430 MUF = 13.83;
fo = 4.8 y h = 505 MUF = 12.83 ;

*Las 02:00*
fo = 3.9 y h = 300 MUF = 14.23;
fo = 4.3 y h = 335 MUF = **14.69** ;
fo = 4.5 y h = 360 MUF = 14.4; Con el método grafico se obtuvo 14.9
Por arriba
fo = 4.6 y h = 430 MUF = 13.53 ;
fo = 4.7 y h = 520 MUF = 12.32;
fo = 4.7 y h = 525 MUF = 12.25 ;

*Las 03:00*
fo = 3.8 y h = 305 MUF = 13.73;
fo = 4.1 y h = 330 MUF = 14.13;
fo = 4.4 y h = 360 MUF = **14,41** ; Con el método grafico se obtuvo 14.3

*Las 04:00*
fo = 3.8 y h = 295 MUF = 14.0 ;
fo = 4.1 y h = 330 MUF = **14.13**;
fo = 4.3 y h = 360 MUF = 14.08; Con el método grafico se obtuvo 14.2

*Las 05:00*
fo = 3.7 y h = 305 MUF = 13.37 ;
fo = 4.0 y h = 335 MUF = **13.66** ;
fo = 4.2 y h = 370 MUF = 13.57; Con el método grafico se obtuvo 13.8

*Las 06:00*
fo = 3.1 y h = 290 MUF = 11.53;
fo = 3.3 y h = 315 MUF = **11.69** ;
fo = 3.5 y h = 350 MUF = 11.66; Con el método grafico se obtuvo 11.8

*Las 07:00*
fo = 5.1 y h = 260 MUF = 20.18 ;
fo = 5.5 y h = 280 MUF = **20.85**  ;
fo = 5.8 y h = 310 MUF = 20.75; Con el método grafico se obtuvo 21.0

***Las 08:00***
fo = 6.9 y h = 265 MUF = 27.00;
fo = 7.5 y h = 285 MUF = 28,16;
fo = 7.9 y h = 310 MUF = **28.26**; Con el método grafico se obtuvo 28.3
Por Abajo
fo = 5.0 y h = 235 MUF = 20.89;
fo = 5.5 y h = 240 MUF = 22.73;
fo = 6.0 y h = 250 MUF = 24,31;

***Las 09:00***
fo = 7.3 y h = 265 MUF = 28.57;
fo = 7.7 y h = 280 MUF = 29.2;
fo = 8.3 y h = 315 MUF = **29.4**; Con el método grafico se obtuvo 29.9
Por Abajo
fo = 5.8 y h = 235 MUF = 24.23;
fo = 6.4 y h = 245 MUF = 26.18;
fo = 6.6 y h = 250 MUF = 26.74;

***Las 10:00***
fo = 8.0 y h = 290 MUF = 29.75;
fo = 8.7 y h = 310 MUF = 31.12;
fo = 9.2 y h = 340 MUF = **31.15**; Con el método grafico se obtuvo 31.5
Por Abajo
fo = 6.2 y h = 265 MUF = 24.26;
fo = 6.6 y h = 270 MUF = 25.55;
fo = 7.0 y h = 275 MUF = 26.81;

***Las 11:00***
fo = 8.0 y h = 280 MUF = 30.33;
fo = 8.6 y h = 300 MUF = **31.39**;
fo = 9.0 y h = 330 MUF = 31.01; Con el método grafico se obtuvo 31.6

***Las 12:00***
fo = 7.7 y h = 295 MUF = 28.36;
fo = 8.2 y h = 320 MUF = **28.78**;
fo = 8.6 y h = 350 MUF = 28.64; Con el método grafico se obtuvo 29.0

***Las 13:00***
fo = 8.3 y h = 315 MUF = 29.40;
fo = 8.9 y h = 335 MUF = 30.40 ;
fo = 9.5 y h = 370 MUF = **30.60**; Con el método grafico se obtuvo 31.0

***Las 14:00***
fo = 8.7 y h = 315 MUF = 30.82;
fo = 9.3 y h = 335 MUF = 31.76;
fo = 9.9 y h = 370 MUF = **31.90**; Con el método grafico se obtuvo 32.0

***Las 15:00***
fo = 9.2 y h = 310 MUF = 32.91;
fo = 10.0 y h = 340 MUF = **33.86**;
fo = 10.5 y h = 375 MUF = 33.56; Con el método grafico se obtuvo 33.9

***Las 16:00***
fo = 9.6 y h = 285 MUF = 36.04;
fo = 10.4 y h = 315 MUF = **36.84**;
fo = 11.0 y h = 350 MUF = 36.63; Con el método grafico se obtuvo 37.0

***Las 17:00***
fo = 9.0 y h = 265 MUF = 35.21;
fo = 9.8 y h = 290 MUF = **36.44** ;
fo = 10.2 y h = 315 MUF = 36.14; Con el método grafico se obtuvo 36.0

***Las 18:00***
fo = 6.9 y h = 265 MUF = 27.00;
fo = 7.5 y h = 290 MUF = **27.89** ;
fo = 7.8 y h = 315 MUF = 27.63; Con el método grafico se obtuvo 28.0

***Las 19:00***
fo = 4.3 y h = 260 MUF = 16.65;
fo = 4.6 y h = 280 MUF = **17.44** ;
fo = 4.8 y h = 310 MUF = 17.17; Con el método grafico se obtuvo 17.5

***Las 20:00***
fo = 3.7 y h = 380 MUF = 11.73;
fo = 4.0 y h = 415 MUF = **12.03** ;
fo = 4.2 y h = 455 MUF = 11.94; Con el método grafico se obtuvo 12.2

***Las 21:00***
fo = 4.1 y h = 350 MUF = 13.65;
fo = 4.4 y h = 380 MUF = 13.95 ;
fo = 4.6 y h = 395 MUF = **14.27**; Con el método grafico se obtuvo 14.5

***Las 22:00***
fo = 4.1 y h = 335 MUF = 14.00 ;
fo = 4.4 y h = 355 MUF =  **14.53**;
fo = 4.6 y h = 385 MUF = 14.48; Con el método grafico se obtuvo 14.6

***Las 23:00***
fo = 4.3 y h = 310 MUF = 15.38;
fo = 4.5 y h = 320 MUF =  15.79;
fo = 4.9 y h = 355 MUF = **16.18**; Con el método grafico se obtuvo 16.4